\documentclass[twocolumn,showpacs,preprintnumbers,amsmath,amssymb,showkeys]{revtex4}
\usepackage{graphicx}% Include figure files
\usepackage{dcolumn}% Align table columns on decimal point
\usepackage{bm}% bold math
\usepackage{pstcol}
\usepackage{color}
\newcommand{\bel}{\begin{equation}\label}
\newcommand{\ee}{\end{equation}}
\newcommand{\beq}{\begin{eqnarray}\label}
\newcommand{\eq}{\end{eqnarray}}
\newcommand{\bit}{\begin{itemize}}
\newcommand{\eit}{\end{itemize}}
\newcommand{\kn}{\kappa_{n}}
\newcommand{\tg}{(\Delta t)_{g}}
\newcommand{\tb}{(\Delta t)_{b}}
\newcommand{\mcap}{\mathcal{A}^{+}_{n}}
\newcommand{\mcan}{\mathcal{A}^{-}_{n}}

\newcommand{\mvol}{\mathcal{V}_{n}}
\newcommand{\dhal}{\mathcal{D}^{+}_{n,\mbox{\scriptsize halo}}}
\newcommand{\dkern}{\mathcal{D}^{+}_{n,\mbox{\scriptsize ker}}}

\newcommand{\lbd}{\boldsymbol{\lambda^{+}}}

\newcommand\sixth{\ensuremath{{\scriptstyle\frac{1}{6}}}}
\newcommand\shalf{\ensuremath{{\scriptstyle\frac{1}{2}}}}

\begin{document}

\preprint{APS: Physical Review Letters}

\title{Cluster formation in complex multi-scale systems}

\author{J. D. Gibbon}

%\author{Second author}%\email{Second.Author@institution.edu}
\affiliation{Department of Mathematics, Imperial College London, 
London SW7 2AZ, UK}%

\author{E. S. Titi}\affiliation{Department of Computer Science and Applied 
Mathematics, Weizmann Institute of Science, P.O. Box 26, Rehovot, 76100 Israel\\
and\\
Department of Mathematics and Department of Mechanical and Aerospace Engineering, 
University of California, Irvine, CA 92697-3875, USA}
\date{\today}% It is always \today, today,
\begin{abstract}
Based on the competition between members of a hierarchy of length scales in complex multi-scale 
systems, it is shown how clustering of active quantities into concentrated sets, like bubbles 
in a Swiss cheese, is a generic property that dominates the intermittent structure. 
The halo-like surfaces of these clusters have scaling exponents lower than that of 
their kernels, which can be as high as the domain dimension. Examples include spots in 
fluid turbulence and droplets in spin-glasses.
\end{abstract}
%%%%%%%%%%%%                              
\maketitle
%%%%%%%%%%%%
It has long been recognized that active quantities in complex systems of many types are not 
distributed evenly across a domain but cluster strongly into irregular bubbles, as in a 
Swiss cheese. The nomenclature, the nature and shape of the bubbles, and the physics in each 
subject is substantially different: spottiness in high Reynolds number fluid turbulence 
\cite{Frischbk,BT49,MS91,Zeff,SB05,MBbk,AJMnotes,RK85,VM94} and boundary layers \cite{Emmons51}; 
droplet formation in spin-glasses \cite{FH,MM1,PY}; clustering behaviour in networks 
\cite{net1,WBE1}; the preferential concentration of inertial particles \cite{EF,SS,Bec,BCCM,HPut} 
with applications to rain initiation by cloud turbulence \cite{FFS};  clustering of luminous 
matter \cite{BakChen1,BakChen2,HughPac,BakPac} and magnetic bubbles in astro-physics 
\cite{magbub}, are some examples. These clusters display strong features whose typical length 
scales are much shorter than their averages, thus 
raising the question of the nature of the interface between them and the surrounding 
longer scale regions. For instance, in spin glasses the `surface' of the droplets has 
a fractal-like structure whereas the droplets themselves have the full domain dimension 
\cite{PY}. In fluid turbulence the concentrated sets on which vorticity accumulates are 
tubes and sheets, although the fractal nature of these is unclear.  
These sets dominate the associated Fourier spectra which display a spikiness that is the 
hallmark of what is usually referred to as intermittency 
\cite{Frischbk,BT49,MS91,Zeff,SB05,MBbk,AJMnotes,RK85,VM94}.
The ubiquity of this irregular bubble-like topology suggests the existence of a set of 
underlying organizing principles in complex multi-scale systems. \textit{Using 
simple but broadly applicable mathematical ideas, this paper will demonstrate that the 
dominant physical principle behind clustering is the existence of a hierarchy of length 
scales whose members are in competition.}

Consider a $d$-dimensional system whose smallest characteristic (integral) scale $L$ 
is such that the system is statistically homogeneous on boxes $\Omega = [0,L]^{d}$. 
Moreover, it is endowed with the following two properties. 
Firstly, at each point $x\in\Omega$, it possesses an ordered set of length 
scales $\ell_{n} = \ell_{n}(x)$ associated with a hierarchy of features 
labelled by $n\geq 2$
\bel{ls1}
L > \ell_{1}\geq \ell_{2} \geq \ldots \geq \ell_{n}\geq \ell_{n+1}\ldots
\ee
%\par\medskip\noindent
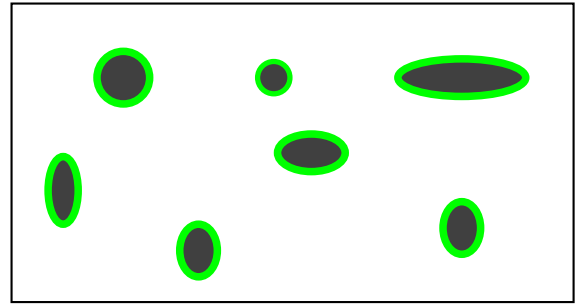
\begin{figure}
\begin{pspicture}(0,0)(8,4)
\psframe(0,0)(7.5,4)
% green = lightgray
\psellipse[linewidth=1mm,
linecolor=green,fillstyle=solid,fillcolor=darkgray](4,2)(.5,.3)
\psellipse[linewidth=1mm,
linecolor=green,fillstyle=solid,fillcolor=darkgray](6,3)(0.9,.3)
\psellipse[linewidth=1mm,
linecolor=green,fillstyle=solid,fillcolor=darkgray](0.7,1.5)(0.25,.5)
\psellipse[linewidth=1mm,
linecolor=green,fillstyle=solid,fillcolor=darkgray](6,1)(.3,.4)
\psellipse[linewidth=1mm,
linecolor=green,fillstyle=solid,fillcolor=darkgray](2.5,.7)(.3,.4)
\pscircle[linewidth=.7mm,
linecolor=green,fillstyle=solid,fillcolor=darkgray](3.5,3){0.25}
\pscircle[linewidth=1mm,
linecolor=green,fillstyle=solid,fillcolor=darkgray](1.5,3){0.4}
\end{pspicture}
\caption{\label{1} {\small An illustrative slice through $\Omega$ 
for one value of $n$: the black kernels are surrounded by green(gray) 
halos. Very small-scale behaviour concentrates on the black \& green(gray) 
regions which constitute the set $\mcap$ ($\mathcal{L}_{n}\kn > 1$). 
The halos have scaling exponents lower than those of the black kernels.}}
\end{figure}
The $\ell_{n}$ could be thought of as an ordered set of correlation or 
coherence lengths; their inverses $\kn(x) = \ell_{n}^{-1}(x)$ clearly obey 
$1< L\kn \leq L\kappa_{n+1}$. The second assumption is that the ensemble 
averages of the $L\kn(x)$ are bounded above by some ordered, positive 
parameters of the system satisfying $1 < R_{n} \leq R_{n+1}$
\bel{ls3}
1 < L\left<\kn\right> \leq R_{n}\,.
\ee
The ensemble average $\left<\cdot\right>$ is a spatial average with respect to 
the Lebesgue measure over $\Omega$. Thus, while the ordering of the $\ell_{n}(x)$ 
must be respected at each point, the $\ell_{n}$ themselves could be quite rough; 
e.g. they  could consist of a series of step functions. If they become very small 
near points $x^{*}$ then they must obey $\ell_{n}> O(r^{d-\varepsilon})$ 
($r=|x-x^{*}|$ and $\varepsilon >0$) so as not to violate (\ref{ls3}).  

Following an idea used in ref. \cite{arma2}, consider the real arbitrary 
parameters $0 < \mu < 1$ and $0 < \alpha < 1$ such that $\mu + \alpha = 1$. 
Use H\"{o}lder's  inequality, $|AB| \leq \frac{1}{p}|A|^{p} 
+ \frac{1}{q}|B|^{q}$ where $\frac{1}{p}+\frac{1}{q}=1$, with $p=\mu^{-1}$ 
and $q= \alpha^{-1}$
\beq{rat1}
\left<\kappa_{n}^{\alpha}\right> \leq\left<\kappa_{n+1}^{\alpha}\right> 
&=& \left<\left(\frac{\kappa_{n+1}}{\kappa_{n}}\right)^{\alpha}\kappa_{n}^{\alpha}
\right>\nonumber\\
&\leq&\left<\left(\frac{\kappa_{n+1}}{\kappa_{n}}\right)^{\alpha/\mu}
\right>^{\mu}\left<\kappa_{n}\right>^{\alpha}\,.
\eq
Re-arranging and factoring out a term $\left<\kappa_{n}^{\alpha}\right>$ gives
\bel{rat2}
\left<\left(\frac{\kappa_{n+1}}{\kappa_{n}}\right)^{\alpha/\mu}\right>
%&\geq& \left(\frac{\left<\kappa_{n}^{\alpha}\right>}
%{\left<\kappa_{n}\right>^{\alpha}}\right)^{1/\mu}\nonumber\\ 
\geq \left<\kappa_{n}^{\alpha}\right> 
\left(\frac{\left<\kappa_{n}^{\alpha}\right>}
{\left<\kappa_{n}\right>}\right)^{\alpha/\mu}\,.
\ee
Lower bounds on the ratio $\left<\kappa_{n}^{\alpha}\right>/\left<\kappa_{n}\right>$ 
can be found from (\ref{ls3}) thereby turning (\ref{rat2}) into
\bel{rat3}
\left<\left(\frac{\kappa_{n+1}}{\kappa_{n}}\right)^{\alpha/\mu}
-\left[(L\kappa_{n})^{\mu}R_{n}^{-1}\right]^{\alpha/\mu}\right>\geq 0\,.
\ee
While it is possible that the integrand in (\ref{rat3}) could be positive
everywhere in $\Omega$, this cannot be assumed; the generic case is that 
the integrand could take either sign\footnote{The word `generic' is being 
used to mean `typically'; of course the extreme case is that the integrand 
is positive, in which case no clusters form at all.}. With the definition 
$\mathcal{L}_{n} = L\,R_{n}^{-1/\mu}$ we have the pair of inequalities
\bel{thm1}
\frac{\kappa_{n+1}}{\kappa_{n}}
\gtrless \left(\mathcal{L}_{n}\kappa_{n}\right)^{\mu},
\ee
for which $\geq$ is valid on regions where the integrand is positive, designated 
as \textit{good regions}, and negative ($<$) on \textit{bad regions}.  The term 
$(\mathcal{L}_{n}\kappa_{n})^{\mu}$ on the right hand side of (\ref{thm1}) 
remarkably contains the arbitrary parameter $\mu$ which lies in the range $0 < \mu 
< 1$. Its existence is important because the ordering in (\ref{ls1}) makes it 
clear from (\ref{thm1}) that everywhere within the bad regions ($<$) there are large 
lower bounds on $\kn$ with exponents containing $1/\mu$
\bel{thm3}
\mathcal{L}_{n}\kn > 1\,,
\hspace{.75cm}\Rightarrow\hspace{.75cm}
L\kn > R_{n}^{1/\mu}\,.
\ee
Let $\mcap$ be the set on which $\mathcal{L}_{n} \kn > 1$ and $\mcan$ the set on which 
$\mathcal{L}_{n} \kn \leq 1$. Then all the bad regions ($<$), designated by the clusters 
of black kernels in Figure 1, lie in $\mcap$. The green/gray halos also lie in $\mcap$, 
and correspond to those parts of the good regions ($\geq$) neighbouring the bad. It is in 
these halos where the lower bound $(\mathcal{L}_{n}\kappa_{n})^{\mu}$ becomes operative. 
The white areas of Figure 1 correspond to $\mcan$ in which the $\kn$ can be randomly 
distributed subject to their ordering in (\ref{ls1}).  It is clear from (\ref{rat3}) that 
the existence and location of the clusters may differ for each $n$. A physical picture 
that displays all clusters for every $n$ would be the union $\mathcal{A}^{+} = \cup\mcap$. 

To show that the volume $\mvol^{+}$ of $\mcap$ comprises a small part of $\Omega$, 
Chebychev's inequality relates the normalized Lebesgue measure $m(\mcap)$ to the integral 
of $L\kn$ over $\mcap$
\bel{meas1}
\int_{\mcap}L\kn\,dm \geq m(\mcap)\,R_{n}^{1/\mu} = L^{-d}\mvol^{+}R_{n}^{1/\mu}\,.
\ee
Together with the relation $\int_{\mcap}L\kn\,dm \leq \left<L\kn\right> \leq R_{n}$
we have
\bel{meas3}
m(\mcap) \leq R_{n}^{-\frac{1}{\mu}+1}\,.
\ee
Hence $m(\mcap)$ is significantly smaller than unity and decreases as $R_{n}$ increases. 
Thus $\mcap$ can fill, at most, a small fraction of $\Omega$.
% Table here
With such sparse information it is difficult to estimate the Hausdorff or the fractal 
dimensions of $\mcap$, but it is still possible to estimate scaling exponents \cite{HP}. 
This entails making a third assumption of self-similarity to estimate the smallest number 
of balls  $\mathcal{N}_{n}^{+}$ of radius $\lbd_{n}$ needed to cover $\mcap$. Defining 
$\lbd_{n}$ as
\bel{lamdef}
(\lbd_{n})^{-1} \equiv k_{n}^{+} = \left<\kn^{p}\right>^{1/p},
%\hspace{1cm}p>1
\ee
for some $p>1$, it is clear that $k_{n}^{+}$ cannot be large enough when $p=1$ because of 
(\ref{ls3}).  However, any value\footnote{As $p\to\infty$, $\left<\kn^{p}\right>^{1/p}
\to\sup_{\Omega}\kn$, which certainly lies within $\mcap$. The $p$-dependence of $k_{n}^{+}$ 
is suppressed.} of $p\gg 1$ will do that makes $k_{n}^{+}$ large enough to be 
a member of $\mcap$. The simplest and worst estimate would be to write
\bel{Ndef1a}
\mathcal{N}_{n}^{+}\sim (L/\lbd_{n})^{d}\,.
\ee
Inequality (\ref{meas3}), however, shows that $\mcap$ occupies only a small fraction of 
$\Omega$. A multiplicative factor of $m(\mcap)$ is introduced thus
\bel{Ndef1}
\mathcal{N}_{n}^{+}\sim m(\mcap)\left(\frac{L}{\lbd_{n}}\right)^{d}
= m(\mcap)\,(Lk_{n}^{+})^{d}\,.
\ee
Instead of using (\ref{meas3}) to estimate $m(\mcap)$, an assumption of self-similar scaling 
is introduced that requires that the change in volume of the balls with respect to $n$ should 
scale as $\mathcal{V}^{+}_{n}$ (the volume of $\mcap$) scales to $L^d$. Thus 
\bel{pref1}
m(\mcap) \sim\frac{\mvol^{+}}{L^{d}} 
\sim \left(\frac{\,\lambda_{n+1}^{+}}{\lbd_{n}}\right)^{d}\,.
\ee
We observe that the definition of the set $\mcap$ in principle involves the length scales 
$L$ and $\lambda_{n}^{+}$, but not overtly $\lambda_{n+1}^{+}$. Yet the good and bad 
sets involve all three scales; $L$, $\lambda_{n}^{+}$ and $\lambda_{n+1}^{+}$. The 
self-similarity assumption (\ref{pref1}) is an assumption about the nature of the set 
$\mcap$ that relates successive length scales $\lambda_{n}^{+}$ and $\lambda_{n+1}^{+}$ 
in an \textit{ad hoc}, yet reasonable, fashion. Using (\ref{pref1}) in (\ref{Ndef1}) we 
have 
\bel{Ndef2}
\mathcal{N}_{n}^{+}\sim \left(\frac{\lambda_{n+1}^{+}}
{\lambda_{n}^{+}}\right)^{d}
\left(\frac{L}{\lbd_{n}}\right)^{d} 
=
\frac{(\mathcal{L}_{n}k_{n}^{+})^{2d}}
{(\mathcal{L}_{n}k_{n+1}^{+})^{d}}\,R_{n}^{d/\mu}\,.
\ee
From these, two estimates for $\mathcal{N}_{n}^{+}$ emerge, one each for the green/gray 
halo and black kernel regions of Figure 1, whose scaling exponents\footnote{Since we 
expect $\mathcal{N}_{n}^{+} \gg 1$, the estimate (\ref{Ndef2}) implies that $L\lambda_{n+1}^{+}
\gg (\lambda_{n}^{+})^{2}$. This is consistent with $\kn > L^{-1}$ as in (\ref{ls1}) but 
technically imposes an additional constraint.} are independent of $p$
\bel{Ndef3}
\mathcal{N}_{n}^{+}\lesssim\left\{
\begin{array}{ll}
\left(\mathcal{L}_{n}k_{n}^{+}\right)^{d(1-\mu)}R_{n}^{d/\mu}
& ~~~~\mbox{\small (green/gray~halo)}\\
\left(\mathcal{L}_{n}k_{n}^{+}\right)^{d}R_{n}^{d/\mu}
& ~~~\mbox{\small (black~kernel)}
\end{array}\right.
\ee
For the former, the $>$ direction of the inequality in (\ref{thm1}) has been used 
together with a simple H\"{o}lder inequality
\bel{Hol}
\left<\kappa^{p(1+\mu)}_{n}\right>^{1/p} \geq
\left<\kappa^{p}_{n}\right>^{(1+\mu)/p} = (k_{n}^{+})^{1+\mu}.
\ee
whereas for the latter $\kappa_{n}\leq \kappa_{n+1}$ has been used. In contrast, 
without any evidence of contraction of volume, the formula corresponding to 
(\ref{Ndef1}) for $\mathcal{N}_{n}^{-}$ is
\bel{Ndef4}
\mathcal{N}_{n}^{-}\sim \left(\frac{L}{\lambda_{n}^{-}}\right)^{d}
= \left(\mathcal{L}_{n}k_{n}^{-}\right)^{d}R_{n}^{d/\mu},
\ee
where $k_{n}^{-}$ satisfies $\mathcal{L}_{n}\kappa_{n}^{-}\leq 1$. The uniform 
scaling exponents in (\ref{Ndef3}) are bounded by
\bel{Ddef}
\dhal \leq d(1-\mu)~~~~~~~~\dkern \leq d
\ee
whereas $\mathcal{D}^{-} = d$ from (\ref{Ndef4}). The coefficients $R_{n}^{d/\mu}$ 
in (\ref{Ndef2}) to (\ref{Ndef4}) reflect the fact that this effect is taking place 
only at length scales smaller than $LR_{n}^{-1/\mu}$. 
\begin{table}
\caption{\label{tab:table1} Summary of conclusions regarding the sets 
$\mathcal{A}_{n}^{\pm}$ and the coloured regions in Figure 1.}
\begin{ruledtabular}
\begin{tabular}{llll}
Figure 1 & black & green(gray) & white\\\hline
Set & $\mcap$ & $\mcap$ & $\mcan$\\
Inequality (\ref{thm1})& $<$ (bad) & $\geq$ (good) & $\geq$ (good)\\
Exponent& $\leq d$ & $\leq d(1-\mu)$ & $=d$\\
\end{tabular}
\end{ruledtabular}
\end{table}
The green/gray halo clearly plays the role of an interface of small but finite thickness 
between the $d$-dimensional (white) outer region and the (black) inner kernel whose 
dimension can be as high as $d$ but could be less. When $\dkern$ saturates its upper 
bound we have
\bel{case1a}
\dhal \leq d(1-\mu) < \dkern = d\,.
\ee
For the green/gray region to have an exponent at least $d-1$ (a surface), $\mu$ would lie in 
the range $0 < \mu \leq 1/d$. Without equations of motion, a numerical experiment would 
be necessary to estimate the $R_{n}$ by finding the maximum value of the ensemble average 
$\left<\kappa_{n}\right>$.  In principle $\mu$ could then be found from numerical estimates 
of $\ell_{n}^{crit}\sim LR_{n}^{-1/\mu}$ within the black kernels although if the $\kn$ 
take very large values there it might not be possible to achieve resolution. $\mu$ itself 
may have upper and lower bounds that are themselves $n$-dependent, as in ref. \cite{arma2}.

We now proceed to discuss some examples. The first ideas on clustering came more than half 
a century ago from Batchelor and Townsend \cite{BT49} who observed intermittent behaviour 
in their high Reynolds number flow experiments, closely followed by observations in 
boundary layers by Emmons \cite{Emmons51}. Batchelor and Townsend called this phenomenon 
`spottiness' and suggested that the energy associated with the small scale components is 
distributed unevenly in space and roughly confined to regions which concomitantly become 
smaller with eddy size \cite{Kuo71}.  Mandelbrot then suggested that these clustered sets 
on which energy dissipation is the greatest might be fractal in nature \cite{MB1}. In 
measurements of the energy dissipation rate in the atmospheric surface layer, Meneveau and 
Sreenivasan interpreted the intermittent nature of their signals in terms of multi-fractals 
\cite{MS91}; a newer generation of experiments measuring intense dissipation in turbulent 
flows have been pursued by Zeff \textit{et al} \cite{Zeff}. Sreenivasan and Bershadskii 
have recently suggested that the clustering of high frequencies in a turbulent signal can 
be characterized by a scaling exponent \cite{SB05}. 

The extremely rapid time evolution of sets of high vorticity or strain in fluid turbulence is 
an important issue; many computations exist showing how these take on the nature of 
quasi-one-dimensional tubes and 
quasi-two-dimensional sheets which have short lifetimes \cite{MBbk,VM94}. An alternative to 
studying the problem in a statistical manner is to include time in the ensemble average 
$\left<\cdot\right>$, in which case the semi-infinite nature of the time-axis suggests a 
different measure might be necessary\footnote{For instance, for a Fokker-Planck equation 
the Gibbs measure would be the most appropriate.}. With specific reference to the Navier-Stokes 
equations, analysis is not advanced enough to deal with the full space-time equations (except 
see ref. \cite{CKN}); conventional methods of analysis use Sobolev norms to $L^2$-average over 
space and remove the pressure \cite{CF,FMRT,MBbk} leaving only time. In ref. \cite{arma2} a 
hierarchy of $\kn$ have been constructed which are comprised of ratios of norms (of derivatives 
of order $n$) and therefore functions of time only; thus the clusters of Figure 1 are merely gaps 
in the time-axis.  It is then necessary to prove that they are finite in width and decreasing 
with increasing Reynolds number. This involves finding bounds on $\mu$. 

The second example is that of the low-temperature phase of spin glasses \cite{SK,PMV}. Our 
conclusions regarding the fundamental role played by the competition between members of a 
hierarchy of length scales is consistent with the observation of ultrametricity in spin-glasses, 
a term that is used to denote the presence of a hierarchy of scales \cite{PMV,PR}. This has been 
observed in computations on the low-temperature spin glass phases of the 
Sherrington-Kirkpatrick \cite{SK,PR,HYD} and Edwards-Anderson models \cite{Stariolo}, as 
well as in dynamic phenomena in complexity \cite{BP1}. The results of this paper, and particularly 
with reference to (\ref{case1a}), are consistent with the droplet theory \cite{FH,MM1,PY} where 
the kernel of the droplet is of full dimension $d$ but its surrounding `surface' has a scaling 
exponent $<d$. In fact, Palassini and Young \cite{PY} have shown numerically that 
$\mathcal{D}^{+}_{\mbox{\scriptsize halo}} = 2.58 \pm 0.02$ when $d=3$ and 
$\mathcal{D}^{+}_{\mbox{\scriptsize halo}} = 2.77\pm 0.02$ when $d=4$.

In conclusion, we have shown that in a system endowed with a competitive hierarchy of 
correlation lengths, a clustering effect ensues in which length scales smaller than a 
critical value, and much smaller that the ensemble average scale, aggregate into small 
intense regions. The kernels of these intense regions are surrounded by halos that have 
scaling exponents smaller that of the domain dimension $d$. We have expectations that 
this idea of competition between scales may be a useful paradigm in explaining the 
behaviour of multi-scale systems.

%\par\vspace{-.25cm}\noindent
\begin{acknowledgments}
We wish to acknowledge discussions with Steve Cowley, Charles Doering, Darryl Holm, Roy Jacobs, 
Robert Kerr, Michael Moore, Maya Paczuski, Andrew Parry, Greg Pavliotis, Jaroslav Stark and 
Christos Vassilicos. J.D.G. would like to thank the 
Isaiah Berlin Foundation for travel support and the hospitality of the Faculty of Mathematics 
and Computer Science of the Weizmann Institute of Science where this work was begun. The 
work of E.S.T. was supported in part by the NSF grant number DMS-0204794, an MAOF Fellowship 
of the Israeli Council of Higher Education, the USA Department of Energy under contract number 
W-7405-ENG-36 and the ASCR Program in Applied Mathematical Sciences.
\end{acknowledgments}

\bibliographystyle{unsrt}

\end{document}